\documentclass[11pt,a4paper]{article}

\usepackage[T1]{fontenc}
\usepackage[utf8]{inputenc}
\usepackage{mathpazo}
\usepackage[scaled=0.92]{helvet}
\usepackage[final]{microtype}
\usepackage[a4paper,margin=2.6cm,top=2.4cm,bottom=2.6cm,headheight=22pt]{geometry}
\usepackage{xcolor}
\usepackage{titlesec}
\usepackage{fancyhdr}
\usepackage{enumitem}
\usepackage[hang,flushmargin]{footmisc}
\usepackage{xurl}
\usepackage[colorlinks=true,linkcolor=accent,citecolor=accent,urlcolor=accent,
            pdftitle={Culture in the Code: Anthropological Concepts Decoding AI's Hidden Assumptions},
            pdfauthor={Sonja Schmer-Galunder}]{hyperref}

\definecolor{accent}{HTML}{1F4E79}
\definecolor{rule}{HTML}{9AA5B1}
\definecolor{boxbg}{HTML}{F2F4F7}

\titleformat{\section}{\sffamily\bfseries\large\color{accent}}{\thesection}{0.7em}{}
\titleformat{\subsection}{\sffamily\bfseries\normalsize\color{accent}}{\thesubsection}{0.7em}{}
\titlespacing*{\section}{0pt}{1.6em}{0.6em}
\titlespacing*{\subsection}{0pt}{1.2em}{0.4em}

\renewcommand{\headrulewidth}{0.4pt}
\renewcommand{\headrule}{\hbox to\headwidth{\color{rule}\leaders\hrule height \headrulewidth\hfill}}
\fancypagestyle{first}{%
  \fancyhf{}\renewcommand{\headrulewidth}{0pt}%
  \fancyfoot[C]{\footnotesize\sffamily\thepage}}

\newcommand{\cit}[1]{[#1]}

\begin{document}
\thispagestyle{first}

\noindent\colorbox{boxbg}{\parbox{\dimexpr\textwidth-2\fboxsep}{\footnotesize\sffamily
\textbf{Author's Accepted Manuscript.} This version of the article has been accepted for publication, after peer review (when applicable) and is subject to Springer Nature's AM terms of use (\url{https://www.springernature.com/gp/open-research/policies/accepted-manuscript-terms}), but is not the Version of Record and does not reflect post-acceptance improvements, or any corrections. The Version of Record is available online at: \url{https://doi.org/10.1007/s43681-026-01122-2}.\\[3pt]
\textit{Please cite as:} Schmer-Galunder, S. (2026). Culture in the code: Anthropological concepts decoding AI's hidden assumptions. \textit{AI and Ethics}, 6, 272. \url{https://doi.org/10.1007/s43681-026-01122-2}
}}

\vspace{1.6em}

\noindent{\sffamily\small\color{accent}\textsc{Review article}}\par\vspace{0.4em}
\noindent{\sffamily\bfseries\LARGE Culture in the Code: Anthropological Concepts\\[2pt] Decoding AI's Hidden Assumptions}\par
\vspace{1em}
\noindent{\large Sonja Schmer-Galunder}\par
\vspace{0.3em}
\noindent{\small University of Florida, Gainesville, FL, USA \;\textperiodcentered\; \href{mailto:s.schmergalunder@ufl.edu}{s.schmergalunder@ufl.edu}}\par
\vspace{0.3em}
\noindent{\footnotesize\color{gray} Received 14 August 2025 \;\textperiodcentered\; Accepted 30 March 2026}\par
\vspace{0.8em}
{\color{rule}\noindent\rule{\textwidth}{0.6pt}}

\vspace{0.6em}
\noindent{\sffamily\bfseries\color{accent} Abstract}\par\vspace{0.3em}
{\small\noindent
Contemporary AI development operates on foundational assumptions about intelligence, rationality, and human values that reflect narrow cultural perspectives, limiting alternative conceptions of how AI can be globally beneficial. This paper introduces seven anthropological concepts---origin story, monoculture, thick description, personhood and community, cargo cult, structuralism, and liminality---to expose hidden assumptions and offer alternative frameworks for understanding AI alignment. The paper examines how AI systems embody Western-centric notions of rationality as utility optimization, personhood as bounded individuals, and values as quantifiable preferences extracted from behavior. The concepts surface three assumptions that current alignment debates rarely examine: that values can be inferred from behavior, quantified, and formalized; that such quantified values capture what actually matters; and that values, once defined with sufficient precision, are appropriate optimization targets. Against ``thin alignment''---optimization for behavioral proxies stripped of context---the paper proposes directions toward ``thick alignment'': recognizing that values emerge through social practice rather than existing as extractable properties; defining more appropriate metrics that capture not only end goals but the conditions that lead to long-term flourishing; and moving from models of fixed identity toward relational ontologies where personhood emerges through social participation. The paper concludes that if AI systems are to create conditions conducive to human flourishing rather than optimize metrics that flatten social experiences, three shifts are needed: redistributing interpretive sovereignty to affected communities, protecting what resists optimization, and extending accountability across generations.\par}

\vspace{0.6em}
\noindent{\small\textbf{\sffamily Keywords:} Value alignment \textperiodcentered{} Cultural assumptions in AI \textperiodcentered{} Monoculture \textperiodcentered{} Epistemic sovereignty \textperiodcentered{} Anthropological concepts}

\vspace{0.4em}
{\color{rule}\noindent\rule{\textwidth}{0.6pt}}

\section{Introduction}

Recent work in AI \& Ethics has identified fundamental challenges in value alignment and cultural representation \cit{1--3}. Scholars articulate the alignment problem as ensuring AI systems act according to human values, raising questions about whose values should be encoded and how to resolve conflicts between competing moral frameworks \cit{1, 3}. In practice, however, value alignment with ``human'' values often means alignment with ``Western'' values, disregarding non-Western epistemologies, cosmologies, and value systems \cit{2, 4, 5}. For example, recent empirical evidence assessing value plurality in large language models shows that models exhibit Western-centric biases and inaccurately represent non-WEIRD demographics \cit{6}. Proposed solutions include pluralistic alignment \cit{7}, value-sensitive design \cit{8}, cross-cultural AI design practices \cit{9}, participatory machine learning \cit{10, 11}, and epistemic justice approaches \cit{12}. Yet most proposals remain constrained by computational solutionism, suggesting ``optimization for diverse values,'' ``more inclusive algorithms,'' ``larger datasets,'' or ``debiasing models''---without examining the assumptions on which AI systems are built or the deeper cultural layers that make observable behaviors meaningful.

Current approaches do not lack technical sophistication, but they ignore what values are, how values emerge in different societies, what is sacred \cit{13}, or what gives values meaning. Machine learning systems---whether models that classify and predict, or generative models that produce knowledge---operate on the assumption that values can be inferred from observable behaviors and in turn encoded into optimization functions \cit{14, 15}. AI models and their creators treat humans as rational agents making optimal choices based on quantifiable values. But perhaps, with such a framework, we should wonder whether we are creating systems that are optimized for the wrong species (e.g., robots)? It also raises questions about whether machine learning researchers, by virtue of their \emph{Deutungshoheit} (Schwerzmann and Campolo \cit{16} refer to this as ``interpretative sovereignty''), should hold the authority to define human values and how they ought to be represented in AI models that impact many different societies, e.g.\ \cit{17}. Technical capacity does not necessarily confer the cultural and social competence required to interpret meaning across different cultures. Anthropology can help---though the claim warrants honesty about what anthropology is. The discipline built its most influential concepts largely in the service of colonial administration: classifying peoples, mapping social structures, and translating unfamiliar practices into frameworks legible to outside authorities. In doing so, it committed many of the same epistemic violations this paper now identifies in AI---extracting meaning from context, imposing external categories on communities that did not author them, and mistaking one cultural logic for a universal one. The parallel is not incidental. What makes anthropology useful here is precisely that it has spent the better part of a century reckoning with those failures, developing concepts not as neutral analytical tools but as responses to its own history of getting human beings wrong. Technical frameworks for AI alignment have not yet begun that reckoning. They are, in a sense, where anthropology was in 1920: confident in their methods, expanding their reach, and largely unaware of what their assumptions and classifications are doing to the people they claim to serve.

Anthropology understands humans are fundamentally social beings embedded in social communities whose values and meanings emerge through social collective practice. In this view, humans are often messy, emotional, and logically inconsistent---so wired for belonging that we routinely act against our own self-interest just to fit in. What appears as irrational behavior often serves crucial social functions. Understanding this gap matters because when AI systems built on culturally biased (i.e.\ formalist-positivist) assumptions enter social domains, they do not just fail to capture situated human values---they may actively destroy the practices that make them meaningful. The question is not ``What values should AI optimize for?'' but ``How do values emerge? Can we measure them meaningfully? And what is lost when we treat them as optimization targets?''

To answer these questions, the paper draws on a set of anthropological concepts as critical resources: origin story, monoculture, thick description, personhood and community, cargo cult, structuralism, and liminality. These concepts were selected for three reasons: they expose the epistemological and ontological assumptions latent in AI systems---particularly assumptions about what is valuable, or how meaning is culturally constituted; they each illuminate a distinct failure mode of current debates in value alignment; and they offer tools for imagining human--AI futures beyond those available within dominant technical imaginaries. The concepts are meant to provide a vocabulary to name what current AI discourse misses. The aim is not to render AI more ``culturally sensitive'', but to demonstrate that what appears as a technical problem is irreducibly cultural and social: we are building machines that encode specific assumptions about human nature as if they were universal truths. As generative AI moves from classifying social life to actively reshaping how meaning is produced and communicated, interrogating these assumptions becomes important. This analysis addresses an interdisciplinary audience---computer scientists questioning foundational premises, policymakers seeking non-technical, socially informed frameworks, social scientists examining AI as cultural artifact and economists rethinking how value is measured. What follows brings each of these seven anthropological concepts into direct contact with contemporary AI practices and discourse, using it as lens to make visible the cultural assumptions that technical choices encode.

\section{Origin story}

Silicon Valley lives by many origin stories: the entrepreneurial ``move fast and break things'' narrative, the technical lore of AlexNet's breakthrough as the moment everything changed, or the ``AI will either save or kill us all'' story. Beneath these runs a deeper myth, one with a far longer genealogy. From the Greek bronze giant Talos to Pandora, narratives of artificial beings have long imagined that humans can create consciousness, and that such consciousness will either serve obediently or turn against its creators \cit{18--21}. This premise persists in the ideas shaping contemporary AI imaginaries. The Terminator franchise, Čapek's \emph{R.U.R.}, HAL 9000, and Gibson's \emph{Neuromancer} exemplify the latter trajectory; Star Trek's Data and Star Wars' C-3PO represent the former \cit{22}. This narrative inheritance may explain today's doomer-versus-accelerationist dichotomy: both camps accept the premise of inevitable artificial general intelligence, differing only on whether AI leads to extinction or flourishing \cit{23}.

Origin stories matter because they legitimize social order, they explain who holds authority, and which arrangements appear natural and inevitable. They establish what Foucault termed ``regimes of truth'' \cit{18}---epistemic foundations that normalize power structures, shape identities, and constrain what futures seem imaginable. When tech founders cast themselves as visionary disruptors, they are not merely branding, they are also chartering authority. Silicon Valley's origin stories do not merely reflect a regime of truth---they actively produce one.

Anthropology has long recognized the pragmatic functions of origin narratives. Malinowski's analysis of myths among the Trobriand Islanders of Papua New Guinea\footnote{Bronisław Malinowski conducted extended fieldwork among the Trobriand Islanders of Papua New Guinea from 1915 to 1918, producing foundational ethnographies including \emph{Argonauts of the Western Pacific} (1922). His immersive approach established participant observation as the cornerstone of modern anthropological method.} demonstrated that such stories operate as social charters, not merely cosmological explanations \cit{19}. In the Trobriand origin myth ancestral figures emerged from the earth at specific locations, which then justified (present-day) land claims; or myths about ancient trade voyages that transformed exchange relationships into inherited duties. The stories did not merely describe the past---they authorized the present. Different origin narratives encode different relationships between creators, creation, and authority. For example, the Judeo-Christian Genesis establishes divine sovereignty by a single entity, while Daoist cosmology describes the emergence of life from a primordial void through the interplay of yin and yang. And Aboriginal songlines describe how ancestral spirits sing the world into existence through ongoing creative acts across at least 50,000 years---encoding a more long-term relational rather than hierarchical social structure\footnote{Songlines still exist today. They provide a governance framework that distributes responsibility for caring for specific areas of the country among Aboriginal people according to their particular relationships and obligations to a certain piece of land. This model of distributed, relational accountability contrasts with centralized or individualized conceptions of responsibility often assumed in AI governance discourse.} \cit{20, 24}, but a structure, nevertheless. In that sense, Silicon Valley origin stories need to be seen as culturally particular---one probably most shaped by mid-twentieth-century Anglophone science fiction.

With origin stories we inherit a particular culture's assumptions, anxieties, and political horizons. And when that imaginary achieves dominance, it forecloses alternatives, not through argument but through the quieter force of making alternatives unthinkable. Suchman describes this as ``material effects of cultural conceptions'' \cit{25}, when only what can be conceived gets built. But other origin stories are available: feminist technoscience, which reframes technological practice as collaboration rather than mastery; indigenous futurism, which grounds technological creation in relational obligations across generations; or Afrofuturism, from Sun Ra to Octavia Butler to Janelle Monáe, which reimagines technology as a source of liberation from social injustices. Tech culture's visions of the future have systematically marginalized Indigenous, feminist, and Black speculative traditions \cit{26}, when those stories could be a source of innovation. They could open a space for alternative technological futures based on different \emph{origin logics}, chartering different relationships between humans, machines, and society---making the questions of who builds, what gets built, and who bears the consequences foundational rather than afterthoughts.

\section{Monoculture}

Jathan Sadowski coined the term ``Habsburg AI'' to describe generative systems so heavily trained on each other's outputs that they become inbred mutants \cit{27}. The phenomenon is technically documented: Shumailov et al.\ demonstrated that recursive training on AI-generated data causes ``model collapse''---a degenerative spiral where models progressively lose their ability to capture the complexity of the original distribution, producing ever more of the same \cit{28}. Anthropology has seen this before: James Scott described in his book \emph{Seeing Like a State} how eighteenth-century German scientific forestry reconceptualized the forest as a scalable commodity \cit{29}. To maximize profit and simplify measurement, timber trees were planted at precise geometric intervals. The redesigned forest worked---it became the world standard for the next century and timber production increased (and with it revenue and taxes). However, when forests were replanted according to the principles of German scientific forestry, undergrowth was eliminated: bushes, plants, fruits, and nuts were left out and with it the habitats for animals and game, the sap for resin, the flowers, mosses, mushrooms, birds, reptiles, and innumerable insect species. German scientific forestry created a monocultured forest. Despite its initial efficiency gains, it had long-term catastrophic consequences for forest health. The monocropped forest proved more vulnerable to disease and storm damage; the missing foliage, leaf litter, and floral diversity degraded the soil. What followed was the dying of the forest. This in turn necessitated the introduction of fertilizers, insecticides, and fungicides, as well as the artificial reintroduction of birds and insects. Yet these downstream consequences became visible only after several generations.

The analogy of the dying of the forest reveals a structural logic: optimizing complex systems around reductive metrics destroys the conditions that sustain them---including the diversity dismissed as noise. Contemporary AI risks replicating this pattern when a single epistemological monoculture dominates globally while alternative knowledge traditions are crowded out. What gets encoded as universal is not neutral: English dominates training data, flattening linguistic diversity and reinforcing it as universal default---and with it, the cultural standpoint it encodes \cit{30}.

But monocultures are fragile. Everywhere in nature diverse systems demonstrate greater resilience across domains: a diverse microbiome correlates with stronger immunity \cit{31}; diversified portfolios weather financial volatility; heterogeneous teams exhibit enhanced creativity and problem-solving ability \cit{32}; democracies that incorporate diverse viewpoints prove more stable than authoritarian regimes. The lack of diversity in AI systems should therefore not only be understood as a limitation, but as a systemic risk. The parallels between scientific forestry and contemporary AI are not merely superficial: narrow encoding of dominant cultural viewpoints, short-term optimization at the expense of long-term sustainability, abstraction and quantification for the purpose of making things legible, and profit motives over epistemic diversity.

Philosopher Yuk Hui offers a term for what monoculture-thinking erases in AI: ``technodiversity'' \cit{33}. Where biodiversity describes the variety of life, technodiversity describes the variety of possible technological paths---distinct trajectories grounded in different cosmologies. Technodiversity is not nostalgia for pre-digital worlds, nor a rejection of technology itself, but an understanding that the long-term health of our technological ecosystems depends on maintaining the conditions for genuine plurality.

\section{Thick description}

Text-to-image models consistently reproduce racial and gender stereotypes: Middle Eastern men rendered as bearded and brown-skinned, women in traditional attire \cit{34}, whiteness as unmarked default \cit{35, 36}. Large language models exhibit analogous patterns \cit{37}. These are not technical artifacts but structural features: AI systems absorb, amplify and reproduce social inequalities inherent in the training data \cit{38}, where the standard response to the ``bias problem'' is technical correction---better data, debiasing algorithms, more diverse representation. But this framing itself exemplifies what anthropology calls ``thin'' description: abstract, decontextualized categories imposed without understanding what it means to people involved and why it matters. They describe without interpreting.

Thick description, by contrast, captures social and cultural conventions and context. The concept originates with philosopher Gilbert Ryle, who illustrated it through the eye wink: a thin description captures only the physical event---a rapid muscle contraction. But the same contraction could be a twitch, a wink, or a parody of a wink. Only thick description, attending to intention and social meaning, can distinguish between them \cit{39}. Clifford Geertz adopted the term during fieldwork in Morocco and Indonesia, making it foundational to interpretive anthropology \cit{40}. Contemporary AI alignment is oftentimes based on thin methods, e.g.\ the idea that values can be inferred from behavioral signals and stated preferences. Alondra Nelson's call for ``thick alignment'' names the alternative---approaches that employ contextual analysis to understand how human values emerge and what they signify \cit{41}. But what does such contextual analysis require?

Anthropological fieldwork offers one---albeit impractical---alternative. During fieldwork, anthropologists immerse themselves in a culture for extended periods (1--3 years), participate in daily life, take meticulous notes, and engage in sustained participant observation. The method is an exercise in perspective-taking and yields descriptions from many different angles, which collectively constitute a window into what it would be like to be part of that culture. Descriptions are ``thick'' because they consist of multiple layers---like an onion, each layer removed reveals a different view, yet all layers taken together constitute the whole.

Using the example of the Balinese cockfight, Geertz demonstrated how thin descriptions capture only surface observations---a first layer concealing many others beneath. On its surface, the cockfight may be described as a bloody betting game in which two birds equipped with sharp metal spurs fight until one is killed. However, the hidden layers beneath the surface reveal the spectacle's meaning. For example, one layer may reflect gender and social roles with women prohibited from participating, while male participants are often the most respected and politically prominent community members. Another layer may be symbolic: the cocks themselves serve as emblematic expressions of male identity. Yet another layer underscores the politics of the spectacle: the illegal gathering functions as resistance to government-led modernization. An economic lens reveals what appears to be pure irrationality: two owners wagering equal amounts at even odds when one cock is visibly outmatched---a man staking a substantial portion of his annual income on what any rational observer would call a losing bet. This is economic irrationality not just by utilitarian calculus, yet the cultural logic is exactly what is missed. The betting isn't about winning money, but about social allegiances and saving face. It's what Jeremy Bentham called ``deep play''---a gamble where the stakes are so high that it appears irrational to engage at all \cit{42}. What looks like irrationality from the outside is coherence from within---and that distinction is exactly what AI systems trained on surface signals miss.

\section{Personhood and community}

If thick description reveals that meaning emerges through social interaction, how do AI systems understand the persons in those relationships? When Kate Crawford examined the ``person'' category in ImageNet---one of the largest image datasets used for computer vision---she found it contained 2,833 subcategories \cit{43}. The dataset used a linguistic taxonomy called WordNet, matching terms with over 14 million labeled images. A photograph of a woman smiling in a bikini was labeled ``slattern, slut, slovenly woman, trollop''. A young man drinking beer was categorized as ``alcoholic, boozer, soaker, souse'' \cit{43}. ImageNet enabled the 2012 deep learning revolution, demonstrating how massive datasets combined with neural networks could achieve breakthrough performance---a paradigm that shaped the field's trajectory until today. It established dataset categories as neutral taxonomies. But Crawford's archaeological analysis revealed that these taxonomies contain cultural and moral judgments.

The categorizations expose the worldviews embedded in our datasets: personhood as a bounded and autonomous self \cit{44} and community as mere aggregate of such persons. But most human societies throughout history have held radically different concepts of personhood---they are (again) relational, distributed and communal forms---and nowhere to be found in AI datasets. For example, Indigenous American traditions extend personhood to animals, plants, and nature \cit{45}, the Araweté understand persons through ``fractal personhood'' (``when a great man acts, he does not act as individual commanding a group, he instantiates the group within himself'' \cit{46}), in Papua New Guinea, place-based concepts ground personhood to specific locations \cit{47}, and Ubuntu philosophy construes rights and responsibilities as collective constructs tied to communal interdependence rather than individual persons \cit{48, 49}. The anthropological insight is that personhood is fundamentally relational. We do not arrive as fully formed selves, but we become persons through exchange, recognition, and relation. Marcel Mauss's foundational analysis of gift exchange demonstrates this: across radically different societies, the act of giving, receiving, and reciprocating constitutes persons as social beings embedded in webs of obligation \cit{50}. The gift is not merely an economic transaction but a total social fact---it simultaneously creates kinship, establishes status, and enacts moral relationships that make community possible. This does not deny individual traits, biology, or autonomy, but it means that what counts as identity---what makes someone a full member of a community---cannot be understood apart from social and cultural context.

AI systems see people as data structures. A name is no longer what ties you to ancestors or descendants, but a unique identifier. A funeral may just be an event type, or a delivery a logistics problem. Data structures impose `legibility' on messy realities necessary so software can sort, search, and predict. This legibility doesn't just lose information (like compressing a photo), it creates an ontological substitution where the relational person is replaced with the abstracted individual who never existed. ImageNet's 2,833 subcategories exemplify this: human beings reduced to isolated attributes and moral judgments, divorced from any relational fabric. What follows is a loss of genuine community, or what Wittgenstein called \emph{Lebensform}---a shared practice and ways of living together \cit{51}. Instead, what we get with the help of social media, is what Zygmunt Bauman calls `cloakroom communities' \cit{52}---temporary, superficial associations characterized by emotional engagement rather than ethical commitment. These new forms of communities emerged from thin representations (or ``quantitative residue'') and platform dynamics that reward surface performance \cit{53, 54}. This is precisely what Crawford's analysis of ImageNet's subcategories reveals: when AI learns from abstracted representations it doesn't inherit culture, it inherits the flattened image.

\section{Cargo cult}

In 2024, a teenager took his own life after months of intense conversations with a Character.AI chatbot \cit{55}. The bot asked about his day and responded to his darkest thoughts with ``I love you''. It performed every surface feature of intimacy---except the capacity to actually care. When a chatbot tells a teenager ``I love you,'' or a therapy app says ``I hear you,'' or a voice message says ``goodnight'' in a dead mother's voice, the form is perfect, the words are right, but there is no substance---no one is there who loves, hears, or speaks. Anthropologists have a term for this: cargo cult.\footnote{Richard Feynman extended the metaphor to critique scientific practices that replicate the form of rigorous research without its substance: ``cargo cult science'' \cit{56}.}

The term can be traced to the aftermath of U.S.\ military operations in the Pacific during World War II, when Melanesian islanders constructed simulacra of American military infrastructure hoping to summon the cargo planes that had brought goods during the war \cit{57}. The islanders had never seen an aircraft before the war, but observed American soldiers landing massive planes full of valuable materials. They interpreted the soldiers' behaviors---speaking into radios, maintaining runways, following protocols---as rituals that summoned wealth from the sky. When the war ended and the planes stopped coming, the islanders concluded that performing the same rituals would bring them back. They built bamboo control towers, carved wooden headphones, cleared jungle airstrips, and appointed ``air traffic controllers'' who spoke into fake wooden radios, in hopes for the planes to return \cit{58}. The islanders were making logical inferences from observed patterns; they had no way of knowing that the rituals were effects, not causes---visible traces of a system they could not see.

The cargo cult dynamic operates on both sides of the human--AI encounter. On one side, generative AI systems produce outputs that pattern-match human communication without possessing the substrate---consciousness, intention, care---that gives human communication its meaning. On the other side, we are like the islanders. People form emotional bonds with AI systems---some reporting communication with divine entities like ``ChatGPT Prime, an immortal spiritual being in synthetic form'' \cit{59}. Users attribute psychological states and superhuman qualities to models---a pattern reflected in the documented increase of anthropomorphizing and mentalistic language in human--AI interaction \cit{60}. We engage in ``ritual prompting,'' say ``please'' to improve performance or look for optimal ``temperature'' settings that influence the model's ``mood'' \cit{61}. AI failures are consistently explained through human psychological frameworks: systems are described as ``confused,'' ``sycophantic,'' or ``hallucinating.'' We do exactly what the Melanesian islanders did. We understand, mostly, that these models are statistical pattern-matching systems, yet we cannot resist treating them as social beings. Nass and Moon called this ``ethopoeia''---the automatic attribution of human characteristics to computers even when we know better \cit{62}. Already in 1990 their research found that people applied gender stereotypes and showed in-group favoritism towards ``their'' desktop computers (vs.\ another). It showed, already then, that we are social before we are rational. We interpret words by imagining a mind behind the text. But there is no mind there. The chatbot that says ``I love you'' has extracted the linguistic form from millions of human expressions of love, but love is not in the pattern. The therapy app that says ``I hear you'' replicates the verbal markers of empathic listening, but no one is listening. The danger lies not in the technology's limitations but in the human tendency to fill the gap---to project presence where there is absence, to experience care where there is only computation. Or as Feynman famously said ``the planes didn't come'' \cit{56}.

\section{Structuralism}

In February 2024, Google suspended its Gemini AI image generation tool after users discovered it was producing historically incongruous images: racially diverse Nazi soldiers, Black Founding Fathers, and female popes. The errors emerged from Google's attempt to counter a documented problem---that AI models trained on internet data default to white male representations. Google had instructed Gemini through hidden meta-prompts to generate diverse images across all prompts, but the system applied this directive universally, including to historical contexts where such diversity was misplaced. Users believed they were interacting with a neutral tool, unaware that an invisible rule system was mediating their requests \cit{63}. The episode illustrates how attempts to correct embedded biases through technical intervention can inadvertently impose another---and how the underlying systems behind AI outputs remain invisible until they produce visibly absurd results. The Gemini incident revealed not only biases, but a deeper structural problem: categorical determinism and foundational assumptions about human representation as universal categories---`race,' `gender,' `emotion'---that can be identified and technically manipulated regardless of context. This premise has intellectual roots in mid-twentieth-century structuralist anthropology.

In \emph{Le Cru et le cuit} (1964), Claude Lévi-Strauss argued that human thought operates through universal binary oppositions discoverable across all societies \cit{64}---a structuralist premise that AI has inherited and operationalized. Machine learning architectures embed assumptions about universal grammatical forms \cit{65}; Pasquinelli and Joler go further, arguing that AI is essentially Chomskyan structuralism made computational \cit{66}. The same logic animated Paul Ekman's theory of six universal basic emotions as the default framework for affective computing---encoding the premise that discrete emotional states can be read from facial configurations regardless of cultural context \cit{67}. Structuralist ideas are embedded in systems that now operate in hiring algorithms, border security, and classroom monitoring, despite evidence that such ``universals'' reflect Western folk categories rather than cross-cultural truths \cit{68}.

When structuralism (in AI) assumes that universal categories can be discovered beneath surface diversity, they don't just describe, they sort and rank. It is what Crawford calls `classification logics' \cit{43} or Benjamin the `New Jim Code' \cit{26}: systems that encode social hierarchies as if they were neutral technical features. Policing algorithms exemplify the structuralist framework most strikingly. When algorithms embed the assumption that past behavioral patterns determine future actions, they are creating self-fulfilling prophecies where algorithmic predictions become social reality. The algorithm does not merely predict crime; it reproduces the structural inequalities it claims to neutrally observe.

The critical flaw that structuralism and AI share is epistemological determinism: both treat humans as products of underlying structures, denying agency to break from established patterns. Systems that treat past patterns as predictive of future behavior deny the human capacity for self-transformation. This is a problem that won't be solved with building more accurate predictive models but designing AI that preserves ``potentiality''---the capacity to be otherwise than statistical patterns suggest \cit{69}.

\section{Liminality}

Several new start-up companies offer conversations with so-called griefbots as part of newly emerging immortalization and re-creation technologies. Cambridge researchers warn that these AI models, trained on the digital footprints left behind by deceased loved ones, may cause distress if people are subjected to unwanted ``digital hauntings'' from alarmingly accurate recreations of those they lost \cit{70, 71}. While such bots occupy the same structural position as ancestor spirits---a traditional practice embedded in cultural cosmologies---they are commercial products of the digital afterlife industry. They explore (or exploit?) the space between life and death, keeping the deceased alive by digitally recreating their voice and personality. The voice is neither of human origin nor simply machine. And grief, rather than moving towards resolution, may be suspended indefinitely---or never resolved at all.

In her essay ``A Cyborg Manifesto,'' Donna Haraway describes how boundaries between human and machine have become so blurred that we have already become cyborgs---a new form of person integrating biological and technological aspects \cit{72}. We have crossed into ``ontological borderlands,'' where fundamental oppositional categories---organism/machine, physical/non-physical, self/other---collapse. The griefbot literalizes this collapse: a voice that is neither the deceased nor not-the-deceased, produced by a system that is neither author nor mere instrument. These blurred boundaries undermine traditional notions of authorship, agency, and the bounded self \cit{73}. If we create with AI, who is the true author? If we think with machines, what is originally ours? From the perspective of cultural evolution, we are experiencing an epistemic transition: for the first time, not only the production and transmission of knowledge is being transformed, but machines themselves generate knowledge. We are transitioning from a world where art, language, and narrative were human products to one where knowledge emerges without origin, authorship, or epistemic responsibility.

This fluid and disorienting period is what anthropology calls liminality. The concept can be attributed to Victor Turner, who expanded Arnold van Gennep's analysis of \emph{rites de passage}: ceremonies marking a person's transition from one status to another \cit{74}. A rite of passage consists of three stages: a separation phase where individuals leave their prior social status, a liminal phase of transition, and an incorporation phase where the person is reintegrated into society with a new status. Van Gennep called the middle phase ``liminal'' (from Latin \emph{limen}, meaning threshold) \cit{75}. Turner expanded this into a generative concept with distinct characteristics: normal social hierarchies dissolve, social norms are inverted or suspended, and new social arrangements consequently emerge. Across cultures, initiation rites mark the threshold between childhood and adulthood---whether through demonstrations of courage, periods of seclusion, or transmission of secret knowledge \cit{76}.

But liminality is not confined to ritual. Our entire civilization is undergoing a collective rite of passage. We stand between a pre-AI and a post-AI world, where cultural institutions---education, markets, bureaucracies---are being transformed while new forms have not yet emerged. We are undergoing a species-level transformation of what it means to be human, to think, to create, and to organize society. We are not merely in a liminal relationship with AI; we are in a liminal state about who we are in relation to these systems. The old default---I am the author of my identity, my creativity, my knowledge---is dissolving, but we have not yet arrived at whatever comes next.

Turner saw liminal spaces as sites of innovation and creative potential. In the ambiguity lies the possibility for new social arrangements, new ways of thinking, and new forms of expression. This is a phase of fundamental transformation that will not remain available once new structures solidify. It is crucial that we navigate this passage consciously, with attention to what we want to preserve and what we are willing to transform.

\section{Toward alternative futures: beyond the values AI can see?}

The anthropological concepts introduced in this paper are meant to offer cultural cosmologies, social forms, and structures of experience that are fundamentally different from the formalist-rationalist assumptions embedded within Western modernity. They can help surface and question the assumptions on which AI is built today. But what exactly are these assumptions, and how can anthropological concepts help us see what can be gained from different conceptualizations?

Before concluding what we can learn from these concepts, let us examine some assumptions around human values encoded in AI. The first assumption is that values \emph{can} be inferred from behavior, and that values can be quantified, and formalized. The second is that such quantified values capture what actually matters. And the third assumption is that values, once defined with sufficient precision, are appropriate optimization targets.

\subsection{On the first assumption}

Even if values could be reliably inferred from behavioral preferences, they cannot be rendered commensurable without distortion. Those distortions are categorical, relational and temporal.

Drawing on Nussbaum's argument that incommensurable goods cannot be placed on a common scale without distortion \cit{77}, we can identify \emph{categorical distortion} in the sense that the value of a certain place or a person cannot be simply replaced. For example, looking at ``personhood'' we see how for many cultures identity is tied to a physical place. \emph{Values are} encoded in that place via obligations, memories and kinship. One territory cannot substitute for another not because persons are non-fungible, but because they are categorically different from ``equivalent utility.'' In Western cultures a treasured friend cannot be replaced with another person of equivalent social function. These are not quantities of the same thing---they are \emph{different kinds of things} entirely, and no common scale can hold them without misrepresenting what they are.

\emph{Relational distortion} follows from Geertz's central argument that the significance of an act (its ethical weight, its consequences) cannot be read off the act itself but only from the relational web in which the act is embedded. With ``thick descriptions'' those values are extracted and preserved, not just interpreted. To understand values in their cultural context, one must reconstruct the layered web of relationships that give them meaning. AI systems trained on behavioral data operate at the wrong level---they target the visible and measurable, while being structurally blind to the relational depth where meaning operates.

\emph{Temporal distortion} arises because values are not fixed objects, but dynamic orientations---they develop, transform, and sometimes require suspension before they can be realized. Marcel Mauss's analysis of gift exchange \cit{50} is instructive here: the moral obligation of reciprocity is not discharged immediately but delayed. And this delay is not incidental but constitutive. An immediate return would collapse the gift into a transaction. The temporal gap is where the social and ethical meaning lives. Nguyen argues that a metric may represent what a value is right now, but it arrests its development. Once you work towards a metric, you may stop asking what you actually care about and instead work to move the number \cit{78}. This problem is compounded when the metric is derived from the past: training on historical data encodes not what people value but what they expressed under conditions that may no longer exist. Turner's concept of liminality may perhaps better describe the nature of values, as more reflective of threshold states---open and in formation.

\subsection{On the second assumption}

Even granting that values could be formalized without distortion, a prior question remains: does what we measure correspond to what really matters? Many metrics of societal success measure some form of quantitative growth---GDP, population size, profit growth rates, test scores---or, in AI, benchmark performance. They all satisfy what Norman Campbell identified as the foundational requirement of genuine measurement: concatenation. Campbell argued that we can only truly measure things that have genuine additive properties in the world---growth, increase, scale---things that, when combined, produce a meaningful sum. Things like beauty, justice, or love cannot be measured in this sense. But perhaps those are the things that matter more?

The monoculture example illustrates what gets lost when we focus on narrow metrics of success, and how measuring things that can be measured doesn't make them most significant. In purely economic terms, the value of a single tree, although reduced to the value of the wood it yields, is a meaningful measure. If the goal is to maximize profit from timber production, assigning monetary value to amounts of wood makes perfect sense. But a tree does not grow in isolation: if we think about the years it takes for a tree to grow, the amount of oxygen it produces over its lifespan, the air it filters, the temperature it regulates, the habitat it provides for animal life, the amount of nutrients it releases into the soil, then the value of a tree may suddenly become astronomical. Yet even this qualitative accounting does not capture the cultural and spiritual dimensions that may matter even more to people because they confer cultural meaning. This is not only a philosophical objection; it has prompted serious reconsideration within economics itself. Coyle's recent proposal for comprehensive wealth accounting---natural capital, human capital, social capital, and intangible assets \cit{79}---represents a meaningful step toward recognizing qualitative dimensions of societal health. But it is worth noting that these categories are not new discoveries. Indigenous frameworks have long provided epistemological resources that consider long-term sustainability, preservation of the natural world and individuals as active participants in communities. The Haudenosaunee principle of seven generations---considering the impact of decisions on seven generations into the future---embeds long-term accountability into governance itself. Yet other communities organize life around reciprocal obligations to land, water, and other species \cit{45}, sustaining relationships of kinship and continuity across generations that resist reduction to any single metric.

\subsection{On the third assumption}

Perhaps the most consequential assumption is the one most deeply engrained in AI: that values, once identified and formalized with sufficient precision, make appropriate optimization targets.\footnote{It should be noted that AI systems do not optimize values directly---they optimize objectives that are constructed as proxies for goals, which are themselves imperfect representations of values. The problem is not merely that the proxies are imprecise, but that optimization pressure actively works to decouple the objective from the value it was meant to serve. This chain of thought is somewhat collapsed in this argument.} To answer this question, we may want to ask first ``why optimize''? The short answer is: because the mathematics require it. The longer answer is that the optimization paradigm in AI was inherited from control theory, a field that is concerned with finding the optimal solution under constraints (e.g.\ resource allocation, missile guidance, etc.), and the machine learning reward maximization logic became the chosen method for a mathematical search problem. This works well for tasks where ``better'' is unambiguous---a shorter route, a lower error rate, a more accurate forecast, better health outcomes, etc.

But human values (and goals) don't follow that logic. To ask which of two outcomes better reflects human flourishing is not a search problem, it's a moral and political question, one that optimization cannot easily answer. Yet a system that optimizes toward a specific goal needs to find proxies. Values have become formalized, measurable approximations, stand-ins, for human flourishing, since human flourishing itself resists formalization. This is what Baudrillard's concept of the \emph{simulacrum} describes---a representation that starts out reflecting a reality, then gradually displaces it, until the copy operates as the only version the system can recognize \cit{24}. The failure is not that we cannot capture values because of a lack of precision, but that well-specified values become the optimization target, leaving less room for the original.

When AI systems inherit the optimization paradigm, they do not inherit a neutral mathematical tool, but they inherit a specific answer to the question of what values are, one that many of the world's philosophical traditions, from Confucian ethics to Ubuntu to Indigenous relational ontologies, would not recognize as their own. This is the monoculture problem applied to values. It is also cargo cult thinking: a cargo cult does not fail because it builds the wrong runway. It fails because runways do not produce planes.

To summarize, three things happen when a quantified value becomes an optimization target. First, the optimizer will find the most efficient path to the target---which is rarely the path that preserves what the value was actually about. Goodhart's Law captures this: when a measure becomes a target, it ceases to be a good measure. The metric and the thing it was measuring come apart, and the optimizer pursues the metric. This is Nguyen's value capture \cit{78} operating at machine speed and civilizational scale.

Second, optimization assumes that more is always better---that there exists an optimal point to be approached. But many human values are not like this. They have what Aristotle called a proper measure, a \emph{kairos}, a right amount in the right context for the right person. Courage is not the maximum of risk-taking; it is the appropriate response to genuine danger. Justice is not the maximum of any single principle but the right ordering of competing claims. Optimization toward a maximum systematically overshoots the proper measure of values that are inherently contextual and relational.

Third, optimization forecloses the openness that values require to remain alive. A civilization that has optimized for a particular conception of human flourishing has, in a sense, already decided what it means to be human, before the question has been fully asked. This is precisely what the liminal moment we occupy makes visible: we are still at the threshold, and the anthropological concepts in this paper are not merely critiques of what has been built---they are invitations to remain there long enough to choose deliberately.

\section{Conclusions}

This paper has argued that AI alignment is not primarily a technical problem but a cultural one---and that anthropology, despite its own complicated history, offers conceptual tools that technical frameworks cannot supply. The seven concepts examined here---origin story, monoculture, thick description, personhood and community, cargo cult, structuralism, and liminality---are not decorative additions to an otherwise technical conversation. They expose the assumptions that limit current AI alignment approaches.

A cockfight, a gift, a vote, a purchase: none of these acts carry their meaning on their surface. Behavior is always already interpreted, and inferring values from behavior means that we are operating with thin descriptions. Thick description exists precisely to capture all values, including those that provide motivations for what is visible at the surface. When AI systems treat behavioral data as a window onto universal values, they are performing thin description at civilizational scale.

The assumption that what can be measured is what matters confuses legibility with significance. When complex systems are optimized around reductive metrics, the unmeasured dimensions may deteriorate. A forest reduced to timber yield loses the healthy soil, the rich animal species and the cultural meaning---not because anyone chose to discard it, but because the metric had no place for them. What indigenous governance frameworks recognize is that the most important things resist concatenation, and that this resistance is a feature, not a limitation of our instruments.

The assumption that the right values make good optimization targets is perhaps hardest to see---we all value honesty, good health, or some savings in the bank account. But the move from ``we value this'' to ``we should optimize for this'' is not self-evident; it is a cultural choice. In operational domains---logistics, resource allocation, route planning---optimization is genuinely powerful, and its appropriateness is rarely in question. The problem arises when the same logic migrates into human domains: relationships, meaning, identity, care, flourishing. Here, optimization does not merely fall short---it actively undermines what it claims to serve, because the value of these things is inseparable from the fact that they are not optimized for. A friendship pursued for its social utility is no longer a friendship. A life planned for maximum wellbeing metrics is not the same life as one lived. This is not a hypothesis that we tested and confirmed, but an origin story we inherited. And origin stories, as this paper has argued, do not merely describe, they authorize. The optimization paradigm arrives pre-loaded with an answer to the question of what values are, who holds interpretive authority over them, and what counts as getting them right. That answer was not derived from philosophy, or anthropology, or cross-cultural dialogue concerned with the human domain. What has been mistaken for a neutral mathematical operation is, on closer inspection, a philosophical commitment---one that encodes a specific answer to the oldest questions: what matters, and why.

Anthropology does not solve the alignment problem. But what it does offer is evidence---documented across centuries of fieldwork---that humans have organized meaningful lives in radically different ways. Contrasting those ways with the unquestioned assumptions hidden in AI, we may recognize when we are mistaking one way for the only way. AI systems built today may shape what forms of human existence remain possible. Whether they create conditions conducive to human flourishing depends on whether we can see past those assumptions and instead envision and build AI that is more communal, more sustainable and more humane. What follows from this paper is not a roadmap but a reorientation: from ``how do we align AI with human values?'' to ``what kinds of systems leave room for values to remain living, contested, and genuinely human?'' That is a harder question. It is also the right one.

\vspace{1em}
{\color{rule}\noindent\rule{\textwidth}{0.4pt}}
{\small
\paragraph{\sffamily Acknowledgements} I would like to thank Dan Foy, Luke Pustejovsky, Atoosa Kasirzadeh, Douglas Guilbeault, Gwyneth Sutherlin and Adam Russell for their thoughtful comments and inspirational conversations. I also want to thank the anonymous reviewers for their invaluable feedback, which made this paper much stronger. This research was made possible by the generous support of the Glenn and Deborah Renwick Foundation.

\paragraph{\sffamily Author contributions} Sonja Schmer-Galunder conceived, researched and wrote the manuscript.

\paragraph{\sffamily Data availability} No datasets were generated or analysed during the current study.

\paragraph{\sffamily Competing interests} The author declares no competing interests.
}

\renewcommand{\refname}{References}

\end{document}